# The Study on the Nonlinear Computations of the DQ and DC Methods

Wen Chen (Corresponding author ) and Tingxiu Zhong

Mechanical Engineering Dept., Shanghai Jiao Tong Univ., Shanghai 200030, China

Permanent mail address: P. O. Box 2-19-201, Jiangshu University of Science & Technology, Zhenjiang City, Jiangsu Province 212013, P. R. China

Present mail address (as a JSPS Postdoctoral Research Fellow): Apt.4, West 1$^{st}$ floor, Himawari-so, 316-2, Wakasato-kitaichi, Nagano-city, Nagano-ken, 380-0926, JAPAN

E-mail: chenw@homer.shinshu-u.ac.jp

Present email: chenwwhy@hotmail.com

**Abstract**

This paper points out that the differential quadrature (DQ) and differential cubature (DC) methods due to their global domain property are more efficient for nonlinear problems than the traditional numerical techniques such as finite element and finite difference methods. By introducing the Hadamard product of matrices, we obtain an explicit matrix formulation for the DQ and DC solutions of nonlinear differential and integro-differential equations. Due to its simplicity and flexibility, the present Hadamard product approach makes the DQ and DC methods much easier to be used. Many studies on the Hadamard product can be fully exploited for the DQ and DC nonlinear computations. Furthermore, we first present SJT product of matrix and vector to compute accurately and efficiently the Frechet derivative matrix in the Newton-Raphson method for the solution of the nonlinear formulations. We also propose a simple approach to simplify the DQ or DC formulations for some nonlinear differential operators and thus the computational efficiency of these methods is improved significantly. We give the matrix multiplication formulas to compute efficiently the weighting coefficient matrices of the DC method. The spherical harmonics are suggested as the test functions in the DC method to handle the nonlinear differential equations occurring in global and hemispheric weather forecasting problems. Some examples are analyzed to demonstrate the simplicity and efficiency of the presented techniques. It is emphasized that innovations presented are applicable to the nonlinear computations of the other numerical methods as well.

**Nomenclature**

$C^{N \times M}$ = The set of N×M real matrices.

$\circ$ = Hadamard product of matrices.

$\otimes$ = The Kronecker (tensor) product of matrices.

$A^{\circ q}$ = The Hadamard power of matrix A, where q is a real number.

$\vec{f}$ = The vector $[f_1, f_2, ..., f_N]^T$.

$\overline{A}_x$, $\overline{A}_y$ = The DQ weighting coefficient matrices, modified by the corresponding boundary conditions, for the 1st order partial derivatives along x- and y- directions, respectively.

$\overline{B}_x$, $\overline{B}_y$ = The DQ weighting coefficient matrices, modified by the corresponding boundary conditions, for the 2nd order partial derivatives along x- and y-directions, respectively.

$\vec{U}$ = The vector $[U_1, U_2, ..., U_N]^T$.

$\{x_j\}$ = The vector $[x_1, x_2, ..., x_N]^T$.

$U_{,x}$, $U_{,xx}$, $U_{,y}$ = The partial derivative of function U(x, y) with respect to variables x and y.

t = Time variable.

x, y = Independent variables.

$\overline{E_x}$, $\overline{E_y}$ = The DC weighting coefficient matrices, modified by the corresponding boundary conditions, for the 1st order derivatives along x- and y- directions, respectively.

$\overline{F_x}$, $\overline{F_y}$, $\overline{F_{xy}}$ = The DC weighting coefficient matrices, modified by the corresponding boundary conditions, for the 2nd order derivatives along x- and y- directions, respectively.

$W$ = The vector $\{W_1, W_2, \cdots, W_N\}^T$.

◊ = The SJT product of matrix and vector.

D{ } = Frechet derivative operator in the Newton-Raphson method.

$I_x$, $I_y$ = The unit matrix corresponding to the number of grid points along the x- and y- directions, respectively.

$\overline{A_{ij}}$, $\overline{B_{ij}}$, $\overline{C_{ij}}$ = The DQ weighting coefficients, modified by the corresponding boundary conditions, for the 1st, 2nd and 3rd order partial derivatives, respectively.

## 1. Introduction

The differential quadrature (DQ) method, introduced by Bellman et al. [1-3], is an easy and efficient numerical method for the rapid solution of various linear and nonlinear differential and integro-differential equations. Other researchers [4-12] also made important contributions to this method and its applications. Recently, Bert, Striz and Wang [13, 14] used harmonic functions instead of polynomials as test functions in the DQ method to handle periodic problems efficiently and circumvent the limitation on the number of grid points in the conventional DQ method based on polynomial test functions. Their study showed that the proper test functions are essential for the computational efficiency and reliability of the DQ method. Striz et. al. [15] gave a domain decomposition technique in applying the DQ method to truss and frame structures successfully. Civan [16] and Malik and Civan [17] proposed the differential cubature (DC) method as a competitive numerical technique for the solution of multi-dimensional differential and integro-differential equations. At present the studies on the DC method are few. But this method is very attractive for practical engineering computations.

This paper is the first of a series of works. We hope to present a new framework in applying the DQ and DC methods to nonlinear problems. First, we analyze the advantages of the DQ and DC methods for the nonlinear problems in comparison to the finite element and finite difference methods as well as other global numerical methods. We also give matrix multiplication formulas to obtain the weighting coefficients of the DC method efficiently. The spherical harmonics are suggested as the test functions in the DC method to handle the nonlinear equations occurring in global and hemispheric weather forecasting problems. Second, by introducing the Hadamard product (H-product) of matrices, the DQ and DC formulations for nonlinear problems can be more easily obtained and have an explicit matrix form. Many studies on the Hadamard product can be fully exploited for the DQ nonlinear analysis [18]. Third, the SJT product of matrix and vector ( SJT is the abbreviate of Shanghai Jiao Tong Univ.) is presented to compute accurately and efficiently the Frechet derivative matrix in Newton-Raphson method for the solution of the nonlinear algebraic matrix equations resulting from the DQ or DC formulations in H-product form. Fourth, by converting some nonlinear differential operators into a combination of a linear operator and a simpler nonlinear operator, the cross nonlinear algebraic terms in the DQ and DC formulation for problems of interest are eliminated or reduced and thus the requirements for virtual memory and computational effort are reduced greatly. We provide some examples to demonstrate these innovations in each section. In the appendix, we discuss briefly the applications of the innovations presented to the finite difference method.

## 2. The DQ and DC methods and their advantages for the nonlinear problems

The DQ and DC methods belong to the global numerical techniques. It is well known that the most distinctive feature of the nonlinear problems is their global correlation in comparison to the linear problems. In other words, the function values far from each other in entire computational domain may have much larger relative affection for nonlinear problems. Of course, for linear problems, there exists similar each other dependence, but that is weaker. Compared with finite element (FE) and finite difference (FD) methods based on local interpolation, the global numerical methods use much more function values to approximate the function value at certain grid point, while the FE and FD methods only use less function information, in which the solution at a grid point is approximated only by using those dependent-variable values at adjacent points. The main shortcomings of the global methods are their computational stability and the applicability for problems with complex geometries. The former can be circumvented by using the proper basis functions for problems of interest, the latter can be overcome by coordinate mappings or multidomain approaches. If the solution for problems of interest is sufficiently smooth and we choose proper basis functions, the global methods produce more reliable and rapidly converging results than finite element and finite difference methods, especially for nonlinear problems. From the resulting formulation standpoint, the FE and FD methods result in a sparsely banded coefficient matrix, while the DQ method generates algebraic equations with full characteristic matrices in one dimensional problems. But for two-dimensional domain problems, Civan and Sliepcevich [18] pointed out that the Jacobian matrix in the resulting DQ algebraic formulation contained many null elements, nearly a half of all elements but much less than in the FD and FE formulations. Unlike the DQ method, the DC method is an absolutely global numerical method for multi-dimensional problems. The DQ method is a special case of the DC method in one dimension. The resulting coefficient matrices in the DC formulation for multi-dimensional problems contain much less null element. Therefore, the DC method may be the most efficient numerical technique for multidimensional nonlinear problems [17]. Some applications [1-6, 17, 20-26] also proved much higher efficiency of the DQ and DC methods for the nonlinear problems than the conventional FE and FD methods. In addition, it should be pointed out that the DQ and DC methods are also generally more effective for the linear problems than the FE and FD methods [7-9, 12-15, 17, 19, 33]. The DC method may be especially efficient for differential equations whose derivatives with respect to one coordinate is not only dependent on the function value at that coordinate direction.

On the other hand, the other global numerical methods (Rayleigh-Ritz, Galerkin, etc.) also have similar inherent efficiency for nonlinear computations. However, these methods require one to select initial trial functions satisfying boundary conditions for problems considered, which is not an easy task for many problems in practice. In addition, these methods need much more strenuous formulation effort. In contrast, the DQ and DC method can easily and exactly satisfy a variety of boundary conditions and require much less formulation effort. Recently, the spectral and pseudo-spectral methods have been extensively used in practical engineering especially for the numerical solutions of fluid dynamic problems. The spectral and pseudo-spectral methods are also belong to global numerical methods and are efficient for many linear and nonlinear problems. As was pointed out in reference [10, 13, 22, 27], the DQ-type method is basically equivalent to the collocation ( pseudo-spectral ) methods. But the DQ and DC methods directly compute function values at grid points rather than spectral variables. Thus, they are more explicit and simple for some practical applications. Moreover, compared with the present DQ and DC method, the pseudo-spectral methods require heavy and complicated formulation effort and lack the ease of implementation of the DQ and DC methods [27]. For some cases given in reference [8], the DQ method have somewhat faster convergence rate.

The proper basic functions can improve the convergence rate and reliability of the DQ method [13, 14]. But It should be emphasized that the choice of the basic functions in the DQ and DC methods is different from the Rayleigh-Ritz, Galerkin methods. The DQ and DC methods need not consider the boundary conditions of certain problems to choose the test function. Thus, the effort to determine the basic functions is very little. Generally, we can categorize the problems into several sorts. For example, the harmonic functions are chosen as the test functions for periodic problems, the family of Bessel functions for buckling, deflection and vibration of circular and annular plates. The DQ and DC methods using the polynomial basic functions are usually effective and reliable for a variety of problems if the roots of Chebyshev, Gauss or Legendre polynomials are adopted as the grid spacing. The idea using split range polynomial expansions as the basic functions proposed by Mansell et al. [27] is applicable to the DQ and DC methods for dealing with problems involving steep gradients and discontinuities. For the definite basic functions, the DQ and DC weighting coefficients for certain grid spacing need be computed only once and are independent of any special problems. Therefore, these weighting coefficients can be used repeatedly for various problems.

## 2.1. The DQ method

The essence of the DQ method is that the partial derivative of a function with respect to a variable is approximated by a weighted sum of function values at all discrete points in that direction. Considering a function f(x) with N discrete grid points [7], we have

$$\frac{\partial^m f(x)}{\partial x^m}\bigg|_{x_i} = \sum_{j=1}^{N} w_{ij}^{(m)} f(x_j) \qquad i = 1, 2, \ldots, N. \qquad (2.1\text{-}1)$$

where $x_j$ are the discrete points in the variable domain. $f(x_j)$ and $w_{ij}^{(m)}$ are the function values at these points and the related weighting coefficients, respectively. In order to determine the weighting coefficients $w_{ij}$, equation (2.1-1) must be exact for all polynomials of degree less than or equal to (N-1). To avoid the ill-conditioning the Vandermonde matrix, the Lagrange interpolation basic functions [10, 13, 26, 28] are used as the test functions and explicit formulas for computing the DQ weighting coefficients are obtained. Using a similar procedure, we can obtain the DQ method based on the harmonic functions [13, 14]. The weighting coefficients for high order derivatives can also be obtained by matrix multiplication [4, 5, 7]. For example, considering the DQ weighting coefficients $A_{ij}$, $B_{ij}$, $C_{ij}$ and $D_{ij}$ for the first, second, third and fourth order derivatives, we have

$$B_{ij} = \sum_{k=1}^{N} A_{ik} A_{kj}, \qquad C_{ij} = \sum_{k=1}^{N} A_{ik} B_{kj}, \qquad D_{ij} = \sum_{k=1}^{N} B_{ik} B_{kj},$$
$$i, j = 1, 2, \ldots, N. \tag{2.1-2}$$

## 2.2. The differential cubature method

The DC method is different from the DQ method in that a partial derivative of the function with respect to a coordinate direction is expressed as a weighted sum of the function values at all discrete points in the entire multi-dimensional solution domain rather than simply in that coordinate direction. Considering a two-variable function $f(x, y)$ [16, 17], one obtains the partial derivatives with respect to x and y expressed as

$$\left. \frac{\partial^m f(x, y)}{\partial x^m} \right|_i = \sum_{j=1}^{N} w_{ij}^{(xm)} f_j, \qquad i = 1, 2, \ldots, N. \tag{2.2-1}$$

and

$$\left. \frac{\partial^p f(x, y)}{\partial y^p} \right|_k = \sum_{j=1}^{N} w_{kj}^{(yp)} f_j, \qquad k = 1, 2, \ldots, N. \tag{2.2-2}$$

where j (or i, k) denotes the one-dimensional indexing of the two-dimensional grid points and $f_j$ is the function value at the corresponding grid point. $w_{ij}^{(xm)}$ and $w_{ij}^{(yp)}$ are the DC weighting coefficients for the related partial derivatives. Note that the number of grid points N in the Eqs. (2.2-1) and (2.2-2) is in the entire multi-dimensional domain rather than only in a coordinate direction as in Eq. (2.1-1) of the DQ method. The DC method degenerates into the DQ method in one dimensional problems. Similar to Eq. (2.1-2) of the DQ method, it is straightforward that there exists the following formulas for the DC weighting coefficient matrix $E_x$, $E_y$, $F_x$, $F_x$ and $F_{xy}$ for the 1st and 2nd order derivatives, i.e.,

$$F_x = E_x^2, \qquad F_y = E_y^2, \qquad F_{xy} = E_x E_y = F_y E_x \qquad (2.2\text{-}3)$$

For higher order derivatives, similar formulas exist. As was pointed out by Malik and Civan [17], the major time-consuming calculations in the DC method are to compute the weighting coefficients. The present formulas can effectively reduce these efforts.

Like the DQ method [14], the proper basic functions are important to apply the DC method to certain problems of interest. Malik and Civan [17] used the monomials as the test functions in the DC method for general problems. Other possible basic functions may be more efficient for certain problems. The nonlinear differential equations occurring in global and hemispheric weather prediction are traditionally computed by the spectral method based on the spherical harmonics [29, 30]. A combination of the DQ and DC methods may yield a potential alternative way of numerical modeling for these problems. This may be a very significant application of the DQ and DC methods in practice. The spherical harmonics can be chosen as the test functions in the DC method for partial derivatives with respect to variables along the spherical surface coordinate directions. The partial derivative with respect to variables along the vertical direction can be computed by the DQ method using the roots of Chebyshev, Gauss or Legendre polynomials as the grid points [10, 13].

## 3. Hadamard Product of matrices

For completeness, we briefly state the notation of Hadamard product and power of matrices and some known results on its properties without proofs [18, 31]. Horn [18] provided an excellent survey on the Hadamard product.

**Definition 1.** Let matrices $A=[a_{ij}]$ and $B=[b_{ij}] \in C^{N \times M}$, the Hadamard product of matrices is defined as $A \circ B = [a_{ij} b_{ij}] \in C^{N \times M}$, where $C^{M \times M}$ denotes the set of $N \times M$ real matrices.

**Definition 2**: If matrix $A=[a_{ij}] \in C^{N \times M}$, then $A^{\circ q}=[a_{ij}^q] \in C^{N \times M}$ is defined as the Hadamard power of matrix A, where q is a real number. Especially, if $a_{ij} \neq 0$, $A^{\circ(-1)}=[1/a_{ij}] \in C^{N \times M}$ is defined as the Hadamard inverse of matrix A. $A^{\circ 0}=11$ is defined as the matrix in which all elements are equal to unity.

**Theorem 3.1**: letting A, B and $C \in C^{N \times M}$, then

1> $A \circ B = B \circ A$ \hfill (3-1a)

2> $k(A \circ B)=(kA) \circ B$, where k is a scalar. \hfill (3-1b)

3> $(A+B) \circ C = A \circ C + B \circ C$ \hfill (3-1c)

4> $A \circ B = E_N^T (A \otimes B) E_M$, where matrix $E_N$ (or $E_M$) is defined as $E_N =[\ e_1 \otimes e_1 \vdots \cdots \vdots e_N \otimes e_N]$, $e_i=[0 \cdots 0 \underset{i}{1} \ 0 \cdots 0]$, $i=1, \cdots, N$, $E_N^T$ is the transpose matrix of $E_N$. $\otimes$ denotes the

   Kronecker product of matrices [28, 31]. \hfill (3-1d)

5> if $A \geq 0$, $B \geq 0$, then $A \circ B \geq 0$. \hfill (3-1e)

6> if A>0, B>0, then A∘B>0. (3-1f)

7> If A and B are non-negative, then

$$\lambda_{\min}(A)\min\{b_{ii}\} \leq \lambda_j(A \circ B) \leq \lambda_{\max}(A)\max\{b_{ii}\}$$, where $\lambda$ is the eigenvalue. (3-1g)

8> (detA)(detB)≤det(A∘B), where det( ) denotes the determinant. (3-1h)

9> $$\frac{d(A \circ B)}{dx} = \left(\frac{dA}{dx}\right) \circ B + A \circ \left(\frac{dB}{dx}\right)$$ (3-1i)

Some of the above theorems may not be used in this paper. But since the studies involved with the Hadamard product are few in the literature and these properties are significant for further applications of the Hadamard product to nonlinear computations, we still list them.

## 4. Hadamard product in the DQ or DC formulations for the nonlinear differential equations

In this paper, we assume that the related boundary conditions for all given examples have been applied to the DQ and DC weighting coefficient matrices using the approach proposed in references [13, 17, 33] or these boundary conditions are substituted directly into the DQ and DC weighting coefficient matrices [7, 19]. Therefore, the boundary conditions are no longer considered separately. The DQ weighting coefficient matrices for multi-dimensional problems are stacked into matrix form similar to those for one dimensional problems. It should be noted that the modified DQ weighting coefficient matrices $\overline{A_x}$, $\overline{A_y}$, $\overline{B_x}$ and $\overline{B_y}$ for multi-dimensional problems here are different from the $A_x$, $A_y$, $B_x$ and $B_y$ defined in references [13, 17, 33] in that the present ones are the resulting and stacked DQ weighting coefficient matrices in a multi-dimensional domain, while the latter are the DQ weighting coefficient matrices only in one dimensional sense. The DQ and DC methods are usually used to approximate the partial derivative with respect to spatial variables. The DQ and DC formulations containing derivatives with respect to time t can be solved as initial-value problems by the well-developed techniques such as the Adams-Moulton method, etc. [10, 19, 23].

Considering the nonlinear differential operator $\frac{\partial f(x,y)}{\partial x}\frac{\partial^2 f(x,y)}{\partial y^2}$, its DQ formulation can be expressed in Hadamard product form as

$$\frac{\partial f(x,y)}{\partial x}\frac{\partial^2 f(x,y)}{\partial y^2} = \left(\overline{A}_x \vec{f}\right) \circ \left(\overline{B}_y \vec{f}\right) \qquad (4\text{-}1)$$

The DQ formulations for some other nonlinear differential operators can be expressed in a similar way.

$$xU_{,x} = \{x_j\} \circ \left(\overline{A}_x U\right) \qquad (4\text{-}2a)$$

$$U_{,x} U_{,y} = \left(\overline{A}_x U\right) \circ \left(\overline{A}_y U\right) \qquad (4\text{-}2b)$$

$$UU_{,x} = U \circ \left(\overline{A}_x U\right) \qquad (4\text{-}2c)$$

$$U_{,xy} U_{,x} U_{,y} = \left(\overline{B}_{xy} U\right) \circ \left(\overline{A}_x U\right) \circ \left(\overline{A}_y U\right) \qquad (4\text{-}2d)$$

where $\overline{B}_{xy}$, $\overline{A}_x$ and $\overline{A}_y$ denote the DQ weighting coefficient matrices for the corresponding partial derivatives, modified by the related boundary conditions, respectively.

**Example 1:** The governing equations for the axisymmetric geometrically nonlinear deflection of thin circular plates are given by [34]

$$y^2 \frac{\partial^2 \varphi}{\partial y^2} = \varphi(y)S(y) + Qy^2 \qquad (4\text{-}3a)$$

$$y^2 \frac{\partial^2 S}{\partial y^2} = -\frac{1}{2}\varphi(y)^2 \qquad (4\text{-}3b)$$

where Q is the external force. In terms of the DQ method using Hadamard product, we have

$$\{y_j^2\} \circ \left(\overline{B}_\varphi \vec{\varphi}\right) = \vec{\varphi} \circ \vec{S} + Q\{y_j^2\} \qquad (4\text{-}4a)$$

$$\{y_j^2\} \circ (\overline{B_s}\vec{S}) = -\frac{1}{2}\vec{\varphi}^{\circ 2} \tag{4-4b}$$

where $\overline{B}_\varphi$ and $\overline{B}_s$ represent the DQ weighting coefficient matrices for the second order derivatives, modified by the corresponding boundary conditions [13, 17, 33], respectively. $\vec{\varphi}$ and $\vec{S}$ are the vectors composed of function φ(y) and S(y) values at interior discrete points.

The advantages of the present formulations are their explicit matrix form and the fact that easily programmable algorithmic expressions are obtained. If using the traditional DQ polynomial approximate expression, the formulations will become more complex.

**Example 2**: For multi-dimensional problems, we have the similar Hadamard DQ and DC formulation. For example, consider the following nonlinear differential equation (example 10 of reference [23] )

$$\frac{\partial^2 W(x,y)}{\partial x^2} + \frac{\partial W(x,y)}{\partial x}\frac{\partial^2 W(x,y)}{\partial y^2} = e^{-xy}[xy^2 - 2y + x^3 e^{-xy} - x^4 y e^{-xy}] \tag{4-5}$$

The DC formulation in matrix form for Eq. (4-5) is

$$\phi(\vec{W}) = \overline{F_x}\vec{W} + (\overline{F_x}\vec{W}) \circ (\overline{F_y}\vec{W}) - \{e^{-x_j y_j}[x_j y_j^2 - 2y_j + x_j^3 e^{-x_j y_j} - x_j^4 y_j e^{-x_j y_j}]\} = 0 \tag{4-6}$$

In the following, we deduce the formulation for general quadratic nonlinear partial differential equations in form of ordinary and Kronecker product of matrices. According to the fourth equation (Eq. (3-1d) ) of theorem 3.1, we have

$$(\overline{A_x}\vec{U}) \circ (\overline{A_y}\vec{U}) = E_n^T[(\overline{A_x}\vec{U}) \otimes (\overline{A_y}\vec{U})]E_1 = E_n^T(\overline{A_x} \otimes \overline{A_y})(\vec{U} \otimes \vec{U}) \tag{4-7}$$

where $E_1 = 1$, $(\overline{A}_x\vec{U}) \otimes (\overline{A}_y\vec{U}) = (\overline{A}_x \otimes \overline{A}_y)(\vec{U} \otimes \vec{U})$ [31, 32]. n is the number of interior grid points. Consider the following differential equation for the quadratic nonlinear problems in two-dimensional domain:

$$\sum_{k,l=1}^{P} \alpha_{ij} \frac{\partial^{(k+l)}U}{\partial x^k \partial y^l} + \sum_{\substack{i,j=1 \\ k,l=1}}^{M} \beta_{kl} \frac{\partial^{(i+j)}U}{\partial x^i \partial y^j} \frac{\partial^{(k+l)}U}{\partial x^k \partial y^l} + C = 0 \qquad (4\text{-}8)$$

where c is constant, $\alpha_{ij}$ and $\beta_{ij}$ constant coefficients. We can give the DQ or DC formulation in matrix form for the nonlinear equation (4-8), namely:

$$\sum_{k,l=1}^{P} \alpha_{kl} w^{(k+l)} \vec{U} + E_n^T \left[ \sum_{\substack{i,j=1 \\ k,l=1}}^{M} \beta_{kl} w^{(i+j)} \otimes w^{(k+l)} \right] (\vec{U} \otimes \vec{U}) + C = 0 \qquad (4\text{-}9)$$

where $w^{(k+l)}$ and $w^{(i+j)}$ represent the DQ or DC weighting coefficient matrices, modified by the related boundary conditions, for the corresponding derivatives. The first term in the above formulation is linear and the second term nonlinear. Let

$$\begin{aligned} L_{n \times n} &= \sum_{k,l=1}^{P} \alpha_{kl} w^{(k+l)} \\ Q_{n \times n^2} &= E_n^T \sum_{\substack{i,j=1 \\ k,l=1}}^{M} \beta_{kl} w^{(i+j)} \otimes w^{(k+l)} \end{aligned} \qquad (4\text{-}10)$$

where $L \in C^{n \times n}$ and $Q \in C^{n \times n^2}$. Thus, the formulation (4-9) can be restated as

$$L\vec{U} + Q(\vec{U} \otimes \vec{U}) + C = 0 \qquad (4\text{-}11)$$

The above formulations are explicit and easy to be used. It is straightforward that there exists similar matrix formulations for higher nonlinear problems. It is still possible to simplify the formulation (4-9) further. The further research is under consideration.

## 5. SJT product of matrix and vector

The Newton-Raphson method is a standard numerical technique to solve the nonlinear equation set resulting from the DQ and DC (or the other numerical methods) formulations for the nonlinear differential or integro-differential equations. one of the major time-consuming calculation in the Newton-Raphson method is to compute the Frechet derivative matrix. In this section, we will provide an efficient and explicit procedure to compute the Frechet derivative matrix of the nonlinear DQ and DC formulation in Hadamard product form. First, we herein present a new multiplication operation – SJT product of matrix and vector:

**Definition 3:** If matrix $A=[a_{ij}]\in C^{N\times M}$, vector $B=\{b_j\}\in C^{N\times 1}$, then $A\lozenge B=[a_{ij}\, b_j]\in C^{N\times M}$ is defined as the postmultiplying SJT product of matrix A and vector B, where $\lozenge$ represents the SJT product. If M=1, $A\lozenge B=A\circ B$.

**Definition 4:** If matrix $A=[a_{ij}]\in C^{N\times M}$, vector $B=\{b_j\}\in C^{M\times 1}$, then $B^T\lozenge A=[a_{ij}\, b_i]\in C^{N\times M}$ is defined as the SJT premultiplying product of matrix A and vector B.

Let $\varphi(\vec{U}) = \{\varphi_1(\vec{U})\ \ \varphi_2(\vec{U})\ \ \cdots\ \ \varphi_m(\vec{U})\}^T$, then the Frechet derivative matrix in the Newton-Raphson method for the nonlinear algebraic equations involving $\varphi(\vec{U})$ is defined as

$$D\{\varphi\} = \begin{bmatrix} \dfrac{\partial\varphi_1}{\partial U_1} & \dfrac{\partial\varphi_1}{\partial U_2} & \cdots & \dfrac{\partial\varphi_1}{\partial U_m} \\ \dfrac{\partial\varphi_2}{\partial U_1} & \dfrac{\partial\varphi_2}{\partial U_2} & \cdots & \dfrac{\partial\varphi_2}{\partial U_m} \\ \vdots & \vdots & \vdots & \vdots \\ \dfrac{\partial\varphi_m}{\partial U_1} & \dfrac{\partial\varphi_m}{\partial U_2} & \cdots & \dfrac{\partial\varphi_m}{\partial U_m} \end{bmatrix}. \tag{5-1}$$

Considering the DQ formulation for the nonlinear differential operator in Eq. (4-1) of section 4, its Frechet derivative matrix in the Newton-Raphson method can be obtained by

$$D\{(A_x f) \circ (B_y f)\} = A_x \lozenge (B_y f) + B_y \lozenge (A_x f) \tag{5-2}$$

Eq. (5-2) gives the accurate solutions for the Frechet derivative matrix in problems considered through simple algebraic computations and thus computational effort is reduced greatly. The SJT premultiplying product are related to the Frechet derivative matrices for the DQ formulations such as $\frac{dU^m}{dx} = \vec{A}\vec{U}^m$, i.e.,

$$D\{\vec{A}_x U^m\} = \left(mU^{\circ(m-1)}\right)^T \Diamond \vec{A}_x \tag{5-3}$$

In the following, we give some formulas to compute the Frechet derivative matrices of the DQ and DC formulations for some nonlinear differential operators often encountered in practice:

(1) For $x^q U_{,x} = \{x_j^q\} \circ (\vec{A}_x U)$, where q is real number, one has

$$D\{\{x_j^q\} \circ (\vec{A}_x U)\} = \vec{A}_x \Diamond \{x_j^q\}. \tag{5-4a}$$

(2) For $(U_{,x})^N = (\vec{A}_x U)^{\circ N}$, where N is a real number, one has

$$D\{(\vec{A}_x U)^{\circ N}\} = N\vec{A}_x \Diamond (\vec{A}_x U)^{\circ(N-1)}. \tag{5-4b}$$

For example, $D\{(\vec{A}_x U)^{\circ 2}\} = 2\vec{A}_x \Diamond (\vec{A}_x U)$, $D\{(\vec{A}_x \vec{U})^{\circ(-1)}\} = (-\vec{A}_x) \Diamond (\vec{A}_x \vec{U})^{\circ(-2)}$

$D\{U^{\circ(-1)}\} = (-I) \Diamond U^{\circ(-2)}$ and $D\{U\} = I$.

(3) For $U_{,xy} U_{,x} U_{,y} = (\vec{B}_{xy} U) \circ (\vec{A}_x U) \circ (\vec{A}_y U)$, one has

$$D\{(\vec{B}_{xy}\vec{U}) \circ (\vec{A}_x \vec{U}) \circ (\vec{A}_y \vec{U})\} = ((\vec{A}_x \vec{U}) \circ (\vec{A}_y \vec{U})) \Diamond \vec{B}_{xy} + ((\vec{B}_{xy} \vec{U}) \circ (\vec{A}_y \vec{U})) \Diamond \vec{A}_x \\ + ((\vec{B}_{xy} \vec{U}) \circ (\vec{A}_x \vec{U})) \Diamond \vec{A}_y \tag{5-4c}$$

From the above formulas, we can find that using the SJT product, the Frechet derivative matrix for the DQ or DC formulation of a nonlinear operator can be obtained in the chain rules similar to those for the derivative of a function. The computational effort for a SJT product is only $n^2$ multiplications, which may be the minimum computing effort for Frechet derivative matrix. Thus, the SJT product approach is especially significant for real time nonlinear problems. In addition, the SJT product approach is also easily implemented in a parallel treatment way. The SJT product has the following properties:

If matrix A, B $\in C^{N \times M}$, vector C, D $\in C^{N \times 1}$, then

1. $k(A \Diamond C) = (kA) \Diamond C = A \Diamond (kC)$, where k is a scalar. (5-5a)

2. $(A+B) \Diamond C = A \Diamond C + B \Diamond C$ (5-5b)

3. $A \Diamond C \Diamond D = A \Diamond (C \Diamond D) = (A \Diamond C) \Diamond D$ (5-5c)

4. $(A \circ B) \Diamond D = A \circ (B \Diamond D)$ (5-5d)

Further research in this field needs be conducted in the future.

The Frechet derivative matrix for the DC formulation given in Eq. (4-6) can be expressed as

$$D\{\phi(W)\} = \overline{F_x} + \overline{F_y} \Diamond (\overline{E_x} W) + \overline{E_x} \Diamond (\overline{F_y} W) \qquad (5-6)$$

The nonlinear simultaneous partial differential equations can be decoupled easily by using the Hadamard product and power approach. For example, considering equations (4-4a) and (4-4b),

from Eq. (4-4a) we have

$$S = \bar{\varphi}^{\circ(-1)} \circ \{y_j^2\} \circ (\overline{B_\varphi} \bar{\varphi}) - Q\{y_j^2\} \circ \bar{\phi}^{\circ(-1)} \qquad (5-7)$$

Substituting Eq. (5-8) into Eq. (4-4b), we have

$$\psi(\bar{\varphi}) = \{y_j^2\} \circ \left( \overline{B_s} \left( \bar{\varphi}^{\circ(-1)} \circ \{y_j^2\} \circ (\overline{B_\varphi} \bar{\varphi}) - Q\{y_j^2\} \circ \bar{\varphi}^{\circ(-1)} \right) \right) + \frac{1}{2} \bar{\varphi}^{\circ 2} = 0 \qquad (5-8)$$

The Frechet derivative matrix for Eq. (5-8) can be computed by

$$D\{\psi\} = \left\{\bar{B}_s\left[(-I_n \Diamond \bar{\varphi}^{\circ(-2)}) \Diamond (\{y_j^2\} \circ (\bar{B}_\varphi \bar{\varphi})) + \bar{B}_\varphi \Diamond (\bar{\varphi}^{\circ(-1)} \circ \{y_j^2\}) - Q(-I_n \Diamond \bar{\varphi}^{\circ(-2)}) \Diamond \{y_j^2\}\right]\right\}$$
$$\Diamond \{y_j^2\} + I_n \Diamond \bar{\varphi} \quad (5\text{-}9)$$

where $I_n$ is the unit matrix corresponding to the number of interior discrete points. Eqs. (5-8) and (5-9) only include one dependent variable φ, and the present problems is decoupled successfully. The DQ results using the present Hadamard and SJT product agree well with those in reference [22, 34, 35], and are not presented herein for the sake of brevity.

In order to state the presented approaches more clearly, we provide the following two detailed numerical examples (examples 4 and 5 of reference [36] ). The roots of the shifted Chebyshev polynomials are adopted as the grid spacing in the present examples.

**Example 1**: $y'' + \dfrac{1}{y} + \dfrac{y'^2}{y} = 0; \quad y(0) = 1, y(1) = 2$ (5-10)

Considering the boundary conditions, the DQ formulation for the second order derivative of function y can be stated as

$$\begin{Bmatrix} y''_2 \\ y''_3 \\ \vdots \\ y''_{N-2} \\ y''_{N-1} \end{Bmatrix} = \begin{bmatrix} B_{22} & B_{23} & \cdots & B_{2,N-1} \\ B_{22} & B_{33} & \cdots & B_{3,N-1} \\ \vdots & \vdots & \vdots & \vdots \\ B_{N-2,2} & B_{N-2,2} & \cdots & B_{N-2,N-1} \\ B_{N-1,2} & B_{N-1,2} & \cdots & B_{N-1,N-1} \end{bmatrix} \begin{Bmatrix} y_2 \\ y_3 \\ \vdots \\ y_{N-2} \\ y_{N-1} \end{Bmatrix} + \begin{Bmatrix} B_{21} + 2B_{2,N} \\ B_{31} + 2B_{3,N} \\ \vdots \\ B_{N-2,1} + 2B_{N-2,N} \\ B_{N-1,1} + 2B_{N-1,N} \end{Bmatrix}$$ (5-11)

$= \bar{B}\bar{y} + \vec{b}$

Similarly, one has
$$\bar{y}' = \bar{A}\bar{y} + \bar{a} \quad (5\text{-}12)$$

The formulation for the differential equations in the Hadamard product form is stated as
$$\psi(\bar{y}) = \bar{y} \circ (\bar{B}\bar{y} + \bar{b}) + 1 + (\bar{A}\bar{y} + \bar{a})^{\circ 2} = 0 \quad (5\text{-}13)$$

The iterative formula of the Newton-Raphson method in solution of the nonlinear algebraic equations is

$$\bar{y}^{(k+1)} = \bar{y}^{(k)} - D^{-1}\left(\bar{y}^{(k)}\right) \tag{5-14}$$

where D is the Frechet derivative matrix and can be calculated by using the SJT product, namely,

$$D\{\psi(\bar{y})\} = I \Diamond (B\bar{y} + \bar{b}) + B \Diamond \bar{y} + 2\bar{A}\Diamond(A\bar{y} + \bar{a}) \tag{5-15}$$

The solutions for the linear differential equation

$$y'' = 0; \quad y(0)=1, y(1)=2 \tag{5-16}$$

are adopted as the initial guess values of the present Newton-Raphson iterative scheme. One obtains the convergence results with four iterations. The convergence criteria is the maximum absolute residual of Eq. (5-13) is less than or equal to $10^{-10}$. The maximum relative error of the DQ results is less than 0.001 when using six grid points. Eqs. (5-13) and (5-15) give explicit matrix form for the nonlinear computation of this example. Thus, the application of the DQ method is simplified. It should be noticed that the computational effort for each SJT product in Eq. (5-15) is only $n^2$ multiplications.

In order to handle the following example, we introduce the concept of the Hadamard matrix function, i.e.,

**Definition 5.** If matrix $A=[a_{ij}] \in C^{N \times M}$, then the Hadamard matrix function $f^{\circ}(A)$ is defined as $f^{\circ}(A) = [f(a_{ij})] \in C^{N \times M}$.

**Example 2.** $y'' + \sin(y') + 1 = 0; \quad y(0)=0, y(1)=1 \tag{5-17}$

The DQ formulation for this differential equation can be expressed in the Hadamard matrix function form

$$\psi(\bar{y}) = \bar{B}\bar{y} + \bar{b} + \sin{}^\circ(\bar{A}\bar{y} + \bar{a}) + 1 = 0 \tag{5-18}$$

where $\bar{A}$, $\bar{B}$, $\bar{a}$ and $\bar{b}$ are obtained in the way similar to the above example. The Frechet derivative matrix can be computed by

$$D\{\psi(\bar{y})\} = \bar{B}\bar{y} + \bar{A}\Diamond \cos{}^\circ(\bar{A}\bar{y} + \bar{a}) \tag{5-19}$$

The solutions for linear differential equation

$$y'' + 1 = 0; \quad y(0) = 0, y(1) = 1 \tag{5-20}$$

are chosen as the initial guess for the iteration. We obtain convergence results no more than four iterations when the maximum absolute residual of equations (5-18) is less than to $10^{-10}$. The results in two successive iterates agree to six significant digits.

As is shown in these two examples, the DQ method is very easy to be used for nonlinear problems by using the present Hadamard matrix function, Hadamard product, and SJT product approaches. For very complex problems, it is straightforward that the presented approaches can be used. Moreover, the more complex the problems, the more effective the present approaches. We applied the DQ method using the Hadamard and SJT product to a highly nonlinear sets of partial differential equations of nonlinear bending of orthotropic rectangular plates, and the results are coincident with those by Bert et al. [24]. The governing equations for this case consists of two second order and one fourth order simultaneous nonlinear partial differential equations. The DQ formulations in Hadamard product and power form are decoupled in the way similar to the Eqs. (5-7) and (5-8). Thus, the number of nonlinear algebraic equations are reduced from 3N×3N to N×N, In addition, the SJT product approach also decreases the computational effort for Frechet derivative matrix greatly.

## 6. Approach for simplifying some nonlinear DQ and DC formulations

Following the idea of Bert, Wang and Striz [13], we deduce the DQ method from Lagrangian interpolation formula.

Let $W^2(x) = \sum_{j=1}^{N} \phi_j(x) W_j^2$, (6-1)

where $W_j = W(x_j)$, $\phi_j(x)$ are the Lagrangian interpolation basic functions.

$$(W_i^2)' = \frac{dW^2}{dx}\bigg|_{x_i} = \sum_{j=1}^{N} \frac{\partial \phi_j(x_j)}{\partial x} W_j^2 = \sum_{j=1}^{N} A_{ij} W_j^2 \quad (6-2)$$

where $A_{ij}$ are the DQ weighting coefficients for the 1st order derivatives. Thus,

$$W_i W_i' = \frac{1}{2} \sum_{j=1}^{N} A_{ij} W_j^2 \quad (6-3)$$

In matrix form, we have

$$\{W_j W_j'\} = \frac{1}{2}[A]\{W_j^2\} \quad (6-4)$$

Therefore, the nonlinear term $ww'$ can be approximated by the DQ weighted sum of the square of function values at all discrete grid points. The conventional DQ approximation expression for $ww'$ is

$$W_i W_i' = \{\vec{W} \circ (A\vec{W})\}_i = W_i \sum_{j=1}^{N} A_{ij} W_j \quad (6-5)$$

Obviously, Eq. (6-3) is a simpler formulation than Eq. (6-5) due to the elimination of cross nonlinear algebraic term $(W_i W_j (i \neq j))$. Following this idea, we have

$$\{W_j^2 W'\} = \frac{1}{3}[A]\{W_j^3\} \quad (6-6)$$

$$\{W_j'^2 + W_j W_j''\} = \frac{1}{2}[B]\{W_j^2\} \quad (6-7)$$

where [B] denotes the DQ weighting coefficient matrix for the 2nd order derivatives. Generally, if the nonlinear operator NL{ } can be expressed as

$$NL\{\} = L\{\phi\{\}\} \quad (6-8)$$

where L{ } is a linear operator, and φ{ } is a nonlinear operator but simpler than NL{ }. The DQ or DC approximation for operator NL{ } can be given by

$$NL\{W(x)\}_i = \sum_{j=1}^{N} Q_{ij} \phi(W_j) \quad (6\text{-}9)$$

where $Q_{ij}$'s are the DQ or DC weighting coefficients for linear operator L{ }, which can been obtained by the sum of weighting coefficients of all single derivatives in linear operator L{ }. Since the nonlinear operator φ{ } is simpler than NL{ }, the DQ or DC formulation for operator φ{ } is simpler and easier to be handled than that for operator NL{ }. But it should be noted that the boundary conditions in the present technique are converted into those containing only operator φ{W} before applying the approach proposed by references [13, 17, 33] to modify the DQ weighting coefficient matrices. This conversion procedure is usually easy to be finished. If not using the approach in references [13, 17, 33], the present DQ and DC methods can be used like the conventional DQ and DC method, but the formulations are simplified greatly.

**Example 1**. Consider the Burger's equation [2]:

$$U_{,t} + UU_{,x} = \varepsilon U_{,xx}, \quad \varepsilon > 0 \quad (6\text{-}10)$$

Using the formula (6-3), the DQ formulation for the Eq. (6-10) can be expressed as

$$\frac{dU_i}{dt} + \frac{1}{2}\sum_{j=2}^{N-1} A_{ij} U_j^2 = \varepsilon \sum_{j=2}^{N-1} B_{ij} U_j \quad (6\text{-}11)$$

The conventional DQ formulation given by Bellman [2] is

$$\frac{dU_i}{dt} + U_i \sum_{j=2}^{N-1} A_{ij} U_j = \varepsilon \sum_{j=2}^{N-1} B_{ij} U_j \quad (6\text{-}12)$$

Eq. (6-12) and Eq. (6-11) have an obvious difference in that the former does not contain the cross nonlinear term $U_i U_j$ ($i \neq j$). Thus, its requirements for memory and computing effort are much less than the conventional DQ method.

**Example 2**. The nonlinear Boussinesq equations [37] for shallow water waves can be stated as:

$$\begin{cases} \dfrac{\partial u}{\partial t} + u \dfrac{\partial u}{\partial x} + g \dfrac{\partial h}{\partial x} + \dfrac{1}{3} H \dfrac{\partial^3 u}{\partial t^2 \partial x} = 0 \\ \dfrac{\partial h}{\partial t} + u \dfrac{\partial h}{\partial x} + h \dfrac{\partial u}{\partial x} = 0 \end{cases} \quad (6\text{-}13)$$

Considering $u \dfrac{\partial u}{\partial x} = \dfrac{1}{2} \dfrac{\partial u^2}{\partial x}$, $u \dfrac{\partial h}{\partial x} + h \dfrac{\partial u}{\partial x} = \dfrac{\partial uh}{\partial x}$, the DQ formulation for Eq. (6-13) can be expressed as

$$\begin{cases} \left\{\dfrac{du_j}{dt}\right\} + \dfrac{1}{2}[\overline{A}_{u^2}]\{u_j^2\} + g[\overline{A}_h]\{h_j\} + \dfrac{1}{3}H\dfrac{d^2}{dt^2}\left([\overline{A}_u]\{u_j\}\right) = 0 \\ \left\{\dfrac{dh_j}{dt}\right\} + [\overline{A}_{uh}]\{u_j h_j\} = 0 \end{cases} \quad (6\text{-}14)$$

where g and H are constants. $[\overline{A}_{u^2}]$, $[\overline{A}_h]$, $[\overline{A}_u]$ and $[\overline{A}_{uh}]$ are the DQ weighting coefficient matrices modified by the boundary conditions of function $u^2$, h, u and uh, respectively.

The aforementioned examples eliminate the cross nonlinear algebraic terms in the formulations. Thus, the requirements of virtual memory and computational effort are reduced greatly. This will be significant in practice, especially for some on-line computations. The following gives a numerical comparison between the conventional approach and the technique presented in this section. Considering the differential equation (example 8 of reference [22] ).

$$U_{,x}^2 + UU_{,xx} + U_{,xx} = 0, \quad U(0) = 0, U(1) = 1, \quad (6\text{-}15)$$

First, using the standard procedure, similar to examples 1 and 2 in section 5, we have

$$\psi(U) = \left(\overline{A}_u U + \overline{\alpha}\right)^{\circ 2} + U \circ \left(\overline{B}_u U + \overline{b}\right) + \overline{B}_u U + \overline{b} = 0 \quad (6\text{-}16)$$

Applying the Newton-Raphson method to solve the above nonlinear matrix equation, one can obtain the Frechet derivative matrix by the following formula:

$$D\{\psi(U)\} = 2\overline{A_u} \Diamond (\overline{A_u}U + \overline{a}) + \overline{B_u}\Diamond U + I\Diamond(\overline{B_u}U + \overline{b}) + \overline{B_u} \qquad (6-17)$$

In contrast, we applying the technique presented to this example. Eq. (6-15) can be simplified as

$$\frac{1}{2}\frac{d^2 U^2}{dx^2} + \frac{d^2 U}{dx^2} = 0 \qquad (6-18)$$

The corresponding boundary conditions are

$$U(0)=0, U(1)=1 \quad and \quad U^2(0)=0, \quad U^2(1)=1 \qquad (6-19)$$

Similar to the examples 1 and 2, the second derivatives of functions $U^2$ and $U$ can be approximated by the DQ method as

$$\frac{d^2 \bar{U}^2}{dx^2} = \overline{B_{u^2}}\bar{U}^{\circ 2} + \overline{b}_{u^2}, \quad and \quad \frac{d^2 \bar{U}}{dx^2} = \overline{B_u}\bar{U} + \overline{b}_u \qquad (6-20)$$

Generally, the modified DQ weighting coefficient matrices for functions $U^2$ and $U$ are different, but for this case, $\overline{B_{u^2}}$ and $\overline{B_u}$ for function $U^2$ and $U$ are the same due to the same boundary conditions by chance. Therefore, we has the DQ formulation for Eq. (6-18).

$$\psi(\bar{U}) = \frac{1}{2}\overline{B_{u^2}}\bar{U}^{\circ 2} + \overline{B_u}\bar{U} + \overline{b}_{u^2} + \overline{b}_u = 0 \qquad (6-21)$$

The Frechet derivative matrix can be given by

$$D\{\psi(\bar{U})\} = \bar{U}^T \Diamond \overline{B_{u^2}} + \overline{B_U} \qquad (6-22)$$

Eqs. (6-21) and (6-22) are obviously simpler than Eqs. (6-16) and (6-17).

The solutions for the linear differential equation

$$U_{,xx} = 0; \qquad U(0)=0, U(1)=1 \qquad (6-23)$$

is adopted as the initial guess values of the Newton-Raphson iterative method. The Frechet derivative matrix is computed by Eqs. (6-17) or (6-22), and the convergences are achieved with four iterations. The six grid points are used for this case. Results are displayed under the column $e_u$ of Table 1. The $e_u$ is defined to be the relative errors in computation, i.e., the ratio of the absolute error to the absolute analytical solutions.

Obviously, the technique presented in this section (Eqs. (6-21) and (6-22) ) obtains more accurate DQ results than the conventional procedure (Eqs. (6-16) and (6-17) ). In addition, the computational effort is also reduced greatly in the present approach.

## 7. Conclusions

The present study aims to investigate the nonlinear computations using the DQ and DC methods. The potential advantages of these methods for the nonlinear problems are clearly pointed out. The principal contributions of this paper are to introduce the Hadamard product of matrices to the nonlinear computations of the DQ and DC methods and to present new SJT product of matrix and vector to compute the Frechet derivative matrix in the Newton-Raphson method accurately and efficiently. The Hadamard product approach allows simpler DQ and DC formulations for nonlinear problems. A simple technique in the DQ and DC methods is presented to eliminate or reduce the cross nonlinear algebraic terms in the resulting DQ or DC formulations for some differential operators and thus computational efficiency of the DQ and DC methods can be improved significantly. The practical applications of the DQ and DC methods to nonlinear problems are still few in comparison to the other numerical methods. The present work may provide a guidance to more studies in this field.

The present study shows that the Hadamard product is a very useful mathematical idea for nonlinear problems, But in fact it seems to be seldom applied in practical engineering. As was pointed out by Horn [18], the Hadamard product is also rarely mentioned in the existing linear algebra texts. The present work provides a significant practical application of the Hadamard product. The Hadamard product may be of vital importance in *"nonlinear algebra"* rather than in the conventional linear algebra.

As was pointed out by Bellman et al. [2], the stability of nonlinear computations in the DQ method is a very difficult problem. The H-product may provides a possible manner to deal with this problem. The research in this field will be discussed in subsequent paper. The title of this paper appears to limit the presented innovations to the DQ and DC methods. But in fact, it should be pointed out that the presented techniques (Hadamard and SJT product, etc. ) can be readily extended to nonlinear computations of the other numerical methods such as the finite element, finite difference, spectral, pseudo-spectral, quadrature element [15], discrete ordinate [27, 38], Rayleigh-Ritz, Galerkin, etc.

**APPENDIX**

**On the finite difference method**

The finite difference (FD) formulation for a partial derivative of function W(x, y) can be expressed in matrix form as

$$\left\{\frac{\partial^m W_j}{\partial x^m}\right\}_{P\times 1} = [D]^{(m)}_{p\times p} \{W_j\}_{P\times 1} \tag{A1}$$

where $[D]_{p\times q}^{(m)}$ denotes a sparse FD coefficient matrix. Applying the Hadamard product, one has

$$\left\{\frac{\partial^m W_j}{\partial x^m}\frac{\partial^n W_j}{\partial y^n}\right\}_{P\times 1} = \left(D_x^{(m)}\vec{W}\right)\circ\left(D_y^{(n)}\vec{W}\right) \tag{A2}$$

where $D_x^{(m)}$ and $D_y^{(n)}$ represent the FD coefficient matrix along the x- and y- directions, respectively. As is shown in Eq. (A2), the Hadamard product can be applied to the nonlinear computation of the finite difference method as in the DQ and DC methods. It is straightforward that the SJT product of matrices are also applicable for the finite difference method.

Similar to the formulas (6-4) in section 6 of this paper, we have

$$\left\{W_j \frac{\partial W_j}{\partial x}\right\} = \left\{\frac{1}{2}\frac{\partial^2 W_j}{\partial x^2}\right\} = \frac{1}{2}D_x^{(1)}\vec{W}^{\circ 2} \tag{A3}$$

Note that Eq. (A3) has eliminated the cross nonlinear algebraic terms in the conventional finite difference formulation for the nonlinear operator $W\frac{\partial W}{\partial x}$. The widespread applications of this idea in the finite difference method are the same as in the DQ and DC methods.

**Table 1.**

**Comparison of the Numerical Results with the Analytical Solutions**

|  | Conventional approach $e_u$ | Present approach $e_u$ |
|---|---|---|
| $x_2$ | 7.16E-4 | 5.93E-8 |
| $x_3$ | 1.37E-4 | 3.60E-7 |
| $x_4$ | 7.76E-5 | 2.36E-7 |
| $x_5$ | 5.25E-5 | 3.30E-7 |